\documentclass[letterpaper, 10 pt, conference]{ieeeconf}  

\IEEEoverridecommandlockouts                              

\usepackage{amsmath} 
\usepackage{amssymb}  
\usepackage{amsfonts}
\usepackage{mathtools} 
\usepackage{fourier}
\usepackage{xcolor}  
\usepackage{subcaption}
\usepackage{tikz}
\usetikzlibrary{calc,backgrounds,fit,shapes.callouts}

\usepackage{amsthm}

\newtheorem{thm}{Theorem}
\newtheorem{defn}{Definition}

\newtheorem{asm}{Assumption}

\newtheorem{remark}{Remark}

\usepackage{ifthen}
\newcommand\suppress[1]{}
\newcommand\switchVersion[3]{\ifthenelse{\boolean{#1}}%
{#2\suppress{#3}}%
{\suppress{#2}{\color{magenta}#3}}}
\newboolean{shortVersion}
\setboolean{shortVersion}{true}
\newcommand{\squeezeupSmall}{\vspace{-2mm}}
\newcommand{\squeezeupMid}{\vspace{-4mm}}

\usepackage{lipsum}
\usepackage{etoolbox}
\newcommand{\tinydisplayskip}{%
  \setlength{\abovedisplayskip}{4pt}%
  \setlength{\belowdisplayskip}{4pt}%
  \setlength{\abovedisplayshortskip}{0pt}%
  \setlength{\belowdisplayshortskip}{0pt}
}

\renewcommand{\phi}{\varphi}
\renewcommand{\epsilon}{\varepsilon}

\newcommand{\maji}[1]{\ensuremath{\mathbb{#1}}}
\newcommand{\bigfun}[1]{\ensuremath{\mathcal{#1}}}
\newcommand{\BBB}{{\mathfrak{B}}}
\newcommand{\UUU}{{\mathcal{U}}}

\newcommand{\R}{{\maji{R}}}
\newcommand{\Rnonneg}{{\maji{R}}_{\geq 0}}
\newcommand{\Rpos}{{\maji{R}}_{> 0}}
\newcommand{\Z}{{\maji{Z}}} 
\newcommand{\Zpos}{{\maji{Z}}_{> 0}}
\newcommand{\N}{\maji{Z}_{\geq 0}}
\newcommand{\C}{C}

\newcommand{\SSS}{{\bigfun{S}}}

\newcommand{\GGG}{\bigfun{G}}
\newcommand{\VVV}{\mathcal{V}}

\newcommand{\fw}{\rightarrow}
\newcommand{\bw}{\leftarrow}

\newcommand{\xto}[1]{\xrightarrow{#1}}

\newcommand{\powerset}{{\ensuremath{\wp}}}
\newcommand{\ens}[1]{\left\{ #1 \right\}}

\newcommand{\Ball}[2]{\mathcal{B}_{#2}(#1)} 
\newcommand{\restr}[2]{{#1}_{|#2}}

\newcommand{\len}{{{\text{len}}}}

\newcommand{\model}{{\mathcal{M}}}

\newcommand{\traj}{\sigma}
\newcommand{\Traj}{\mathit{Traj}}
\newcommand{\FTraj}{\mathit{FTraj}}
\newcommand{\IRun}{\mathit{Run}}
\newcommand{\FRun}{\mathit{FRun}}

\newcommand{\FPlay}{\mathit{FPlay}}
\newcommand{\FPlayi}{\FPlay_1}
\newcommand{\FPlayii}{\FPlay_2}

\newcommand{\control}[2]{{#1}/{#2}}
\newcommand{\ini}{\text{in}}

\DeclareMathOperator{\logic}{right-recursive\ LTL} 
 
\DeclareMathOperator{\Logicbf}{\textbf{Right-recursive LTL}} 
\newcommand{\Coloneqq}{\mathrel{\mathop{::}=}}

\DeclareMathOperator{\Globally}{\Box}
\DeclareMathOperator{\Repeatt}{\Box \Diamond}

\DeclareMathOperator{\Until}{\mathop{\mathit{U}}}

\newcommand{\Dis}[2]{[{#1}]_{#2}} 

\DeclareMathOperator{\Inf}{Inf}
\DeclareMathOperator{\MP}{MP}

\newcommand{\mppg}{mean-payoff parity game}
\newcommand{\mppgs}{\mppg{}s}

\newcommand{\pmay}{{\rho_\exists}}
\newcommand{\pmust}{{\rho_\forall}}

\newcommand{\hlength}{1}
\newcommand{\vlength}{1}
\newcommand{\basehlength}{0.8}
\newcommand{\basevlength}{0.5}
\newcommand{\untilhlength}{1}
\newcommand{\untilvlength}{0.5}
\definecolor{rouge}{rgb}{0.8,0,0}
\definecolor{vert}{rgb}{0,0.8,0}
\definecolor{bleu}{rgb}{0,0,0.8}
\makeatletter
\DeclareRobustCommand{\rvdots}{%
  \vbox{
    \baselineskip4\p@\lineskiplimit\z@
    \kern-\p@
    \hbox{.}\hbox{.}\hbox{.}
  }}
\makeatother

\title{\LARGE \bf
Fast Synthesis for Symbolic Self-triggered Control
under $\Logicbf$ Specifications
}

\author{Sasinee Pruekprasert, Clovis Eberhart, and J\'{e}r\'{e}my Dubut
\thanks{The authors are supported by ERATO HASUO Metamathematics for Systems Design Project No. JPMJER1603, JST.
 S.\ Pruekprasert is also supported by Grant-in-aid No. 21K14191, JSPS.
  J.\ Dubut is also supported by Grant-in-aid No. 19K20215, JSPS.}
\thanks{The authors are with the National Institute of Informatics, Hitotsubashi 2-1-2, Tokyo 101-8430, Japan
        {\small \{sasinee, eberhart, dubut\}@nii.ac.jp}.}%
\thanks{C.\ Eberhart and J.\ Dubut are also affiliated with the Japanese-French Laboratory for Informatics, IRL 3527.}%
}

\begin{document}
{\onecolumn\large 
\noindent\textcopyright\ 2021 IEEE. Personal use of this material is permitted.  Permission from IEEE must be obtained for all other uses, in any current or future media, including reprinting/republishing this material for advertising or promotional purposes, creating new collective works, for resale or redistribution to servers or lists, or reuse of any copyrighted component of this work in other works.}
\newpage
\twocolumn
\maketitle
\thispagestyle{empty}
\pagestyle{empty}


\begin{abstract}
We extend previous work on symbolic self-triggered control for non-deterministic continuous-time nonlinear systems without stability assumptions to a larger class of specifications. Our goal is to synthesise a controller for two objectives: the first one is modelled as a right-recursive LTL formula, and the second one is to ensure that the average communication rate between the controller and the system stays below a given threshold. We translate the control problem to solving a mean-payoff parity game played on a discrete graph. Apart from extending the class of specifications, we propose a heuristic method to shorten the computation time. Finally, we illustrate our results on the example of a navigating nonholonomic robot
with several specifications.
\end{abstract}


\section{Introduction}
Self-triggered control has been increasingly attracting attention from academia and
has proven to be a practical control approach, especially for networked control systems
\cite{HJT2012, Hashimoto2019,SEMGL2019, Liu2020}. 
Unlike conventional periodic control schemes, self-triggered control is a proactive control paradigm with a triggering mechanism that prescribes a time when the control signal has to be updated. 
As sensing and actuation are performed only when needed, this control scheme
 can reduce energy consumption and communications across the network. 
Self-triggered controllers can significantly reduce the communications in nonholonomic robots formation control~\cite{SEMGL2019}
and the energy consumption in leader-follower consensus control of networked multi-agent systems~\cite{Liu2020}.

\begin{figure*}[t]
  \centering
  \scalebox{1}{
    \begin{tikzpicture}[basic/.style={draw,align=center}]
      \node[basic,      anchor=west] (system) at (0,0)                   {System\\$\Sigma$};
      \node[basic,      anchor=west] (model)  at ($(system.east)+(2.2,0)$) {Symbolic\\model $\SSS$};
      \node[basic,rouge,anchor=west] (prune)  at ($(model.east) +(2.0,0)$) {Pruned\\model};
      \node[basic,      anchor=west] (game)   at ($(prune.east) +(2.1,0)$) {MPPG\\$\GGG$};
      \node[basic,rouge,anchor=west] (reach)  at ($(game.east)  +(2.0,0)$) {Subgame};
      \path[->] (system)
                  edge
                  node[above,align=center,scale=0.9,font=\small] {discretisation}
                  node[below] {Section~\ref{section: symbolic control}}
                  (model)
                (model)
                  edge[rouge]
                  node[above,align=center,scale=0.9,font=\small] {heuristic\\pruning}
                  node[below] {Section~\ref{section prune}}
                  (prune)
                (prune)
                  edge
                  node[above,align=center,scale=0.9,font=\small] {translation\\to game}
                  node[below] {Section~\ref{subsec: problem trans}}
                  (game)
                (game)
                  edge[rouge]
                  node[above,align=center,scale=0.9,font=\small] {reachability}
                  node[below] {Section~\ref{section reach}}
                  (reach);
      \node[align=center] (control)  at ($(system)-(0,1.5)$) {controller};
      \node[align=center] (disccon)  at ($(model) -(0,1.5)$) {controller};
      \node[align=center] (prunecon) at ($(prune) -(0,1.5)$) {controller};
      \node[align=center] (strat)    at ($(game)  -(0,1.5)$) {winning\\strategy};
      \node[align=center] (substrat) at ($(reach) -(0,1.5)$) {winning\\strategy};
      \path[->] (substrat) edge node[above] {Remark~\ref{rem:reachability}} (strat)
                       (strat)    edge node[above] {Theorem~\ref{thm:strat_to_control}} (prunecon)
                       (prunecon) edge node[above] {Remark~\ref{rem:pruning}} (disccon)
                       (disccon)  edge node[above] {Theorem~\ref{thm: problem_reduction}} (control);
      \path[->] (reach) edge node[left] {solving} (substrat);
      \path[dashed] (system) edge (control)
                    (model)  edge (disccon)
                    (prune)  edge (prunecon)
                    (game)   edge (strat);
    \end{tikzpicture}
  }
  \switchVersion{shortVersion}{\squeezeupSmall}{}
  \caption{Overview of the proposed process.}
  \label{fig: flow}
  \switchVersion{shortVersion}{\squeezeupMid}{}
\end{figure*}
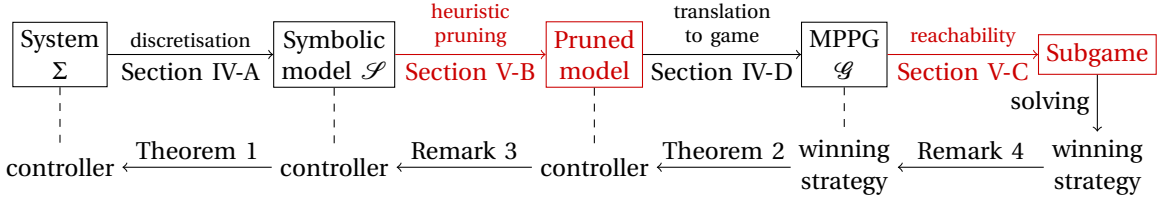

However, previous self-triggered control studies focus on stability \cite{HJT2012, anta2010}, consensus \cite{Liu2020},
and reachability or safety \cite{Hashimoto2019} problems. 
To extend self-triggered control to complex specifications such as temporal logic,
we consider symbolic control,
which is an abstraction-based control approach that constructs a discrete abstraction of the continuous system,
then synthesises a discrete controller that can be refined to a controller for the original system in a sound way \cite{ Zamani2012,liu2016finite, hsu2018lazy, lindemann2019robust, MR2019}.
Using this technique, we can apply the algorithms developed for discrete structures, such as games on graphs, to synthesise provably-correct controllers for complex specifications that can hardly be enforced with conventional control methods.

The first symbolic self-triggered control approach was introduced in~\cite{Hashimoto2019}
for discrete-time deterministic nonlinear systems
under reach-avoid specifications.
It was developed in the symbolic control framework for nonlinear systems without stability assumptions proposed in~\cite{Zamani2012}.
Our previous work in~\cite{icarcv20} extended the concept in \cite{Hashimoto2019} to 
the control of continuous-time non-deterministic nonlinear systems for 
2-LTL specifications: a subclass of Linear Temporal Logic (LTL) specifications 
 strictly more expressive than reach-avoid.
Our control objective was to: 1) control the system to satisfy the given 2-LTL formula and 
2) restrict the limit average control-signal length to stay above a given threshold. 
We reduced the control problem to solving a winning strategy in a mean-payoff parity game played between the controller and the environment.
Then, we transformed the winning strategy into a symbolic self-triggered controller of the original system.

In this work, we extend our self-triggered control methodology proposed in~\cite{icarcv20} to $\logic$ specifications.
We add the temporal operator Until, which is a basic operator in LTL but must be 
treated differently from
the other operators in 2-LTL.
Dealing with Until heavily increases non-determinism in the mean-payoff parity game.
Moreover, 
unlike previous self-triggered control studies that considered constant control signals of different lengths in a sample-and-hold manner
\cite{HJT2012, Hashimoto2019},
 we consider piecewise-constant control signals, resulting in a far larger set of signals choices. 
For these reasons, we develop a heuristic pruning method to speed up the computation.
It disables some control signals based on a notion of reward for execution traces in a B\"uchi automaton corresponding to the $\logic$ formula.
We use the proposed heuristic and reachability analysis, which significantly reduces the computation time.

We study a self-triggered control problem for non-deterministic continuous-time nonlinear systems without stability assumptions,
so the system runs in continuous time (even though the controller only
sees the discrete sequence of states at the end of input signals), so
approaches with discrete-time semantics, such
as~\cite{lindemann2019robust} cannot be applied.
Our approach under-approximates the set of atomic propositions 
that hold along the system trajectories, which is different from \cite{liu2016finite} that considers state abstraction with robust margin.
Moreover, since $\logic$ specifications are strictly more expressive
than  safety and reach-avoid, we cannot use  
feedback- or counterexample-based abstraction refinement techniques
developed for those specifications, 
such as in~\cite{hsu2018lazy}.

Our controller synthesis process is a refinement of the one we
defined in~\cite{icarcv20} (see Fig.~\ref{fig: flow}).
The main steps of the process (in black in Fig.~\ref{fig: flow}) is to
discretise the continuous system $\Sigma$ into a
symbolic model $\SSS_{\eta,\tau,\ell,\mu}$, which is then turned
into a mean-payoff parity game $\GGG_{\Phi,\nu}$.
We then compute a winning strategy for $\GGG_{\Phi,\nu}$, which can be
turned into a controller for $\SSS_{\eta,\tau,\ell,\mu}$, and
ultimately for $\Sigma$.
Since there is currently no algorithm to compute winning strategies
efficiently, we introduce two intermediate steps (in red in
Fig.~\ref{fig: flow}) to reduce the size of the game.
The first one is a heuristic that prunes some transitions of the
symbolic model, and the second one precomputes the reachable part of
the game before solving it.


\textit{Notation:}
 For a vector $x \in \mathbb{R}^m$, we write $\lVert x \rVert$ for its infinity norm $\max\,\{\,\lvert x_i\rvert ~ \mid\, i \in \{1, \ldots, m\}\}$.
Given $x \in \mathbb{R}^m$ and $r \in \mathbb{R}_{> 0}$, we write $\Ball{x}{r}$
for the ball $\{y \in \mathbb{R}^m \,\mid\, \lVert x-y \rVert \leq r\}$ of centre $x$ and radius $r$. 
Finally, given a set $X$, we denote its powerset $\{Y \,\mid\, Y\subseteq X\}$ by $\powerset(X)$.

\section{Control Framework} \label{section: framework} 

\subsection{Non-deterministic Continuous-time Nonlinear System} \label{subsection: system} 
We consider a system modelled by a 6-tuple 
\[
  \Sigma = (X, X_\ini, U, {\UUU}, \xi^\fw, \xi^\bw),
\]
where $X \subseteq \R^n$ and $U \subseteq {\R}^m$ are bounded convex
spaces respectively of states and control inputs,
$X_\ini \subseteq X$ is a set of initial states, 
${\UUU}$ is a set of control signals of the form
$[0,T] \to U$ that assign a control input at all time in $[0,T]$ with $T \in \Rpos$,
and $\xi^\fw$, $\xi^\bw$ $: \R^n \times \UUU \times \Rnonneg \to \powerset(\R^n)$ are forward and backward dynamic functions
such that $\xi_{x,u}^\fw(0) = \xi_{x,u}^\bw(0) = \{x\}$.
We denote by $\len(u)= T$ the length of signal $u:[0,T] \to U$.
The dynamics are defined on $\R^n$, but we are
only interested in the bounded subspace $X$. 

As in \cite{icarcv20}, we require the following assumptions, which
basically ensure that we can bound the distance between points
evolving according to the system dynamics.
Note that this does not imply any stability assumptions on $\Sigma$.

\begin{asm}\label{asm: inc fw and bw complete} 
The system $\Sigma$ is \emph{incrementally forward and backward complete}. Namely, 
$\xi_{x,u}^\fw(t) \neq \emptyset$ and $\xi_{x,u}^\bw(t) \neq \emptyset$ 
for all $(x,u) \in \R^n \times \UUU$ and $t \in [0,\len(u)]$
and, for all $u \in \UUU$,
there are functions $\beta^\fw_u, \beta^\bw_u: \Rnonneg \times \Rnonneg \to \Rnonneg$
such that: 1)  for all $t \in \Rnonneg$, $\beta^\fw_u(\_, t)$ and
$\beta^\bw_u(\_, t)$ are increasing,
and 2)
for all $x_1,x_2 \in \R^n$, $u \in {\UUU}$, and $t \leq \len(u)$,
\begin{itemize}
\item $\forall (x_1',x_2') \in \xi^\fw_{x_1,u}(t) \times \xi^\fw_{x_2,u}(t),
			\lVert x_1' - x_2'\rVert \leq 
      		\beta^\fw_u(\lVert x_1 - x_2 \rVert,t)\rlap{,}$
\item $\forall (x_1',x_2') \in \xi^\bw_{x_1,u}(t) \times \xi^\bw_{x_2,u}(t),
			\lVert x_1' - x_2'\rVert \leq 
        	\beta^\bw_u(\lVert x_1 - x_2 \rVert,t)\rlap{.}$
\end{itemize} 
\end{asm}

\begin{asm}\label{asm: Lipchitz}
For all $u \in \UUU$,
there are functions $\alpha^\fw_u, \alpha^\bw_u: \Rnonneg \times [0, \len(u)] \to \Rnonneg$
such that 1) for all $t \in \Rnonneg$, $\alpha^\fw_u(\_,t)$ and
$\alpha^\bw_u(\_,t)$ are increasing, and 2) for all $x_1, x_2 \in X$ and $t \in [0, \len(u)]$, we have 
\begin{itemize}
\item $\forall y_2 \in \xi_{x_2,u}^\fw(t),
	\rVert x_1 - y_2 \lVert \leq \alpha^\fw_u(\rVert x_1 - x_2 \lVert, t)\rlap{,}
	$
\item $\forall y_2 \in \xi_{x_2,u}^\bw(t),
    \rVert x_1 - y_2 \lVert \leq \alpha^\bw_u(\rVert x_1 - x_2 \lVert, t)\rlap{.}
	$
\end{itemize}
\end{asm}

These functions can typically be computed using Lyapunov functions (see \cite{Zamani2012, Angeli1999, Angeli2002} for details).

\begin{defn}\label{defn: model}
  A \emph{run} (\emph{resp.}\ \emph{finite run}) of $\Sigma$ is a sequence
  $x_0 u_0 x_1 \ldots \in X (\UUU X)^\omega$ (\emph{resp.}\ $x_0 u_0
  \ldots x_k \in X (\UUU X)^*$) such that, for all $i \in \N$ (\emph{resp.}\ $i < k$), 1)
  $x_{i+1} \in \xi^\fw_{x_i,u_i}(\len(u_i))$, 2) $x_i \in
  \xi^\bw_{x_{i+1},u_i}(\len(u_i))$, and 3) $\xi^\fw_{x_i,u_i}(t) \subseteq
  X$ for all $t \leq \len(u_i)$.
  Let $\IRun(\Sigma)$ (\emph{resp.}\ $\FRun(\Sigma)$) denote the set of all
  runs (\emph{resp.}\ finite runs) of $\Sigma$.
\end{defn}
\switchVersion{shortVersion}{\squeezeupSmall}{
\begin{remark}$\model(\Sigma)$ is called a \emph{symbolic model} in~\cite{Zamani2012}.
In this paper, however, we reserve the term \emph{symbolic model} for the discrete-state system in Section  \ref{subsection: symbolic model}.
\end{remark}
}

We need runs of a system to be discrete sequences of states, since
controllers will only observe the state of the system when outputting a
signal, but the system runs in continuous time.
To match discrete runs to continuous sequences of states,
we introduce the notion of trajectory.
\begin{defn}\label{defn: trajectory}
  A \emph{trajectory} of a system $\Sigma$ induced by a run $x_0 u_0 x_1 u_1 x_2 \ldots \in
  \IRun(\Sigma)$ is a function $\traj: \Rnonneg \to X$ such
  that, for all $k \in \N$ and all $t \in \big[\sum_{i < k} \len(u_i), \sum_{i \leq k} \len(u_i)\big]$,
  \switchVersion{shortVersion}{\squeezeupSmall
  \begin{align*}
  \traj&(t) \in \xi_{x_k,u_k}^\fw\big( t - \sum_{i < k} \len(u_i)\big) 
   \cap \xi_{x_{k+1},u_k'}^\bw\big(\sum_{i \leq k} \len(u_i) - t\big),\\
  &\text{where }u_k'(s) = u_k\big(s+t- \displaystyle\sum_{i < k} \len(u_i)\big),  
   \forall s \in [0, \displaystyle\sum_{i \leq k} \len(u_i)-t]. 
  \end{align*} 
  }{
  \begin{equation*}  
   \traj(t) \in \xi_{x_k,u_k}^\fw\big( t - \sum_{i < k} \len(u_i)\big) 
   \cap \xi_{x_{k+1},u_k'}^\bw\big(\sum_{i \leq k} \len(u_i) - t\big),  
  \end{equation*}
  where 
  \begin{equation*} 
  u_k'(s) = u_k\big(s+t- \displaystyle\sum_{i < k} \len(u_i)\big)  
  \text{ for all }
   s \in [0, \displaystyle\sum_{i \leq k} \len(u_i)-t].
  \end{equation*}
  } 
\end{defn}
Let $\Traj(\Sigma, r)$ be the set of trajectories of $\Sigma$ that
are induced by a run $r \in \IRun(\Sigma)$.
For any finite run $r_f \in \FRun(\Sigma)$,
$\FTraj(\Sigma, r_f)$ is the set of finite trajectories defined in
the same way.
Let $\Traj(\Sigma) = \bigcup_{r_0 \in X_\ini} \Traj(\Sigma, r)$.

\subsection{Controlled System} \label{subsection: controlled system}

In this section, we define controllers and controlled systems, and
explain the self-triggered control process.

\begin{defn}\label{defn: model controller}
  A \emph{controller} of $\Sigma$ is a function $\FRun(\Sigma) \to
  \UUU$.
\end{defn}

Technically, a controller is a partial function that only needs to be
defined on the runs it will generate, but we keep this definition for
simplicity.

Let $\control{\C}{\Sigma}$ denote the system $\Sigma$ controlled
under the controller $\C$.
A run $x_0 u_0 x_1 \ldots \in \IRun(\Sigma)$ (\emph{resp.} a finite
run $x_0 u_0 x_1 \ldots x_k \in \FRun(\Sigma)$) is
generated by $\control{\C}{\Sigma}$ if, for all $i \in \N$
(\emph{resp.} $i \in [0, l-1]$), $u_{i} = \C(x_0 u_0 \ldots x_i)$.
Let $\IRun(\control{\C}{\Sigma})$ (\emph{resp.}
$\FRun(\control{\C}{\Sigma})$) denote the set of all runs
(\emph{resp.} finite runs) generated by $\control{\C}{\Sigma}$ from
any $x \in X_\ini$.
The definitions of trajectories $\Traj(\control{C}{\Sigma})$ and
$\FTraj(\control{C}{\Sigma})$ carry over to controlled systems
directly.

Notice that a controller $\C$ of a system $\Sigma$ is defined based on
its runs, rather than its trajectories.
This is because the controller issues control signals based on runs, which only track the states at the end of each signal.

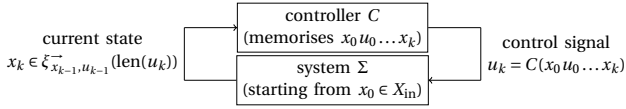
\begin{figure}[t]
      \centering
      \scalebox{0.7}{
        \begin{tikzpicture}
          \node[draw,align=center]                     (controller) at (0,0) {controller $C$\\(memorises $x_0 u_0 \ldots x_k$)};
          \node[draw,align=center,below of=controller] (system)              {system $\Sigma$\\(starting from $x_0 \in X_\ini$)};
          \coordinate (tl) at ($(controller.west)-(1,0)$);
          \coordinate (bl) at ($(system.west)!(tl)!(system.east)$);
          \coordinate (tr) at ($(controller.east)+(1,0)$);
          \coordinate (br) at ($(system.west)!(tr)!(system.east)$);
          \path (controller) edge (tr)
                (tr) edge node[right,align=center] {control signal\\$u_k = C(x_0 u_0 \ldots x_k)$} (br)
                (br) edge[->] (system)
                (system) edge (bl)
                (bl) edge node[left,align=center] {current state\\$x_k \in \xi^\fw_{x_{k-1},u_{k-1}}(\len(u_k))$} (tl)
                (tl) edge[->] (controller);
        \end{tikzpicture}
      }
      \switchVersion{shortVersion}{\squeezeupSmall}{}
      \caption{Overview of the self-triggered control process. The controller issues a control signal based on the previously observed states and issued control signals.}
      \label{fig: control process}
\switchVersion{shortVersion}{\squeezeupMid}{}
\end{figure}
The overview of the control process is illustrated in Fig~\ref{fig: control process}.
First, the controller observes the initial state $x_0 \in X_\ini$ and 
issues control signal $u_0 = \C(x_0)$.  
Then, to reduce the communication rate,
the controller is inactive throughout the duration of $u_0$.
The longer $u_0$ is, 
the less communication is sent across the network.
Since the system is non-deterministic, there are several states that can possibly be reached under the signal $u_0$.
When the signal ends, 
the controller becomes active and resolves the non-determinism by detecting the actual current state $x_1$ 
and issue a new control signal $u_1 = \C(x_0 u_0 x_1)$.
The process is then repeated.

\section{Problem Formulation} \label{section: problem} 

Our goal is to synthesise a controller that satisfies two objectives.
The first one is a temporal specification, described as a $\logic$ formula.
The second one is a communication rate objective, ensuring that the average length of the issued control signals is above a given threshold.


\subsection{Right-recursive LTL Specification} \label{subsection: logic} 

We model the first objective using a fragment of LTL, which we call
$\logic$, and which is also an extension of 2-LTL, studied in~\cite{icarcv20}.
Let $AP$ denote the set of atomic propositions, i.e., assertions that
can be either true or false at each state $x \in X$.
Let $P: X \to \powerset(AP)$ assign the set of atomic propositions
that hold at each state.

\begin{defn}
  Let $\logic$ be the logic whose formulas are the $\Phi$'s generated by the following grammar:
  {\switchVersion{shortVersion}{\tinydisplayskip}{}
  \begin{align*}
    \phi & \Coloneqq \top \mid p \mid  \neg \phi \mid \phi \vee \phi \\
    \Phi & \Coloneqq
    	\phi  \mid
    	\phi \Until \Phi \mid
      \Diamond \phi \mid
      \Box \phi \mid
      \Box\Diamond \phi \mid
      \Diamond\Box \phi \mid
      \Phi \vee \Phi \mid \Phi \wedge \Phi \rlap{,}
  \end{align*}
  }
  where $p\in AP$ is an atomic proposition.
\end{defn}

We call $\phi$'s state formulas and $\Phi$'s path formulas.
A logic specification is written as a path formula.
Here, $\Until$, $\Diamond$, and $\Box$ are given the usual LTL
semantics.
A state $x\in X$ satisfying a state formula $\phi$ is denoted by $x \vDash \phi$. 
We use the same notation $\traj \vDash \Phi$ for a trajectory $\traj: \Rnonneg \to X$ and a path formula $\Phi$.
For all $x \in X$, 
$x \vDash \phi$ is defined as follows: 
{\switchVersion{shortVersion}{\tinydisplayskip}{}
\begin{align*}
	x &\vDash \top &  x &\vDash  p \text{ if } p \in P(x)\\
	x &\vDash \neg\phi \text{ if } x \not\vDash \phi
      &  x &\vDash \phi_1 \vee \phi_2 \text{ if } x \vDash \phi_1 \text{ or }x \vDash \phi_2\rlap{,}
\end{align*}
}
and for all $\traj: \Rnonneg \to X$, $\sigma \vDash \Phi$ is defined as follows:
{\switchVersion{shortVersion}{\tinydisplayskip}{}
\begin{align*}
  \traj &\vDash \phi \text{~~if~~} \traj(0) \vDash \phi\\
  \traj &\vDash \phi \Until \Phi \text{~~if~~} \exists t \in \Rnonneg,\, \restr{\sigma}{t} \vDash \Phi \text{ and } \forall t' \leq t,\, \sigma(t') \vDash \phi
\end{align*}
\begin{align*}
	\traj &\vDash  \Diamond\phi  \text{~~if~~} \exists t \in \Rnonneg,\, \traj(t) \vDash \phi\\
	\traj &\vDash  \Box\phi  \text{~~if~~} \forall t \in \Rnonneg,\, \traj(t) \vDash \phi\\
	\traj &\vDash  \Box\Diamond\phi  \text{~~if~~} \forall t \in \Rnonneg,\, \exists t' > t,\, \traj(t') \vDash \phi\\
  \traj &\vDash \Diamond\Box\phi  \text{~~if~~}\exists t \in \Rnonneg,\, \forall t' > t,\, \traj(t') \vDash \phi\\ 
	\traj &\vDash \Phi_1 \vee \Phi_2  \text{~~if~~}\traj \vDash \Phi_1 \text{ or } \traj \vDash \Phi_2\\
  \traj &\vDash \Phi_1 \wedge \Phi_2  \text{~~if~~} \traj\vDash \Phi_1 \text{ and } \traj \vDash \Phi_2\rlap{,}\\
  \phantom{\traj} & \phantom{\vDash \phi \Until \Phi \text{~~if~~} \exists t \in \Rnonneg,\, \restr{\sigma}{t} \vDash \Phi \text{ and } \forall t' \leq t,\, \sigma(t') \vDash \phi}
\end{align*}
}
\vspace*{-1cm}

\noindent where $\restr{\sigma}{t}(t') = \sigma(t+t')$.
One objective of a controller $\C$ is to control the system in such a way that all  trajectories in $ \Traj(\control{\C}{\Sigma})$ satisfy a given $\logic$ path formula $\Phi$. 
Since $\logic$ extends 2-LTL, it is also more general than the reach-avoid
specifications studied in \cite{Hashimoto2019,MR2019,hashimoto20} (see also Section~\ref{section: illus example}).

\subsection{Controller Synthesis Problem} \label{subsection: problem continuous} 

\begin{defn} \label{defn: problem continuous}
Given a system $\Sigma = (X, X_\ini, U, {\UUU}, \xi^\fw, \xi^\bw)$, a set $AP$ of atomic propositions, a function 
$P: X \to  \powerset(AP)$,
a $\logic$ formula $\Phi$, 
and a threshold $\nu \in \Rpos$,
the \emph{controller synthesis problem} is
to synthesise  a controller $\C:\FRun(\Sigma) \to \UUU$
such that
\begin{itemize}
  \item all finite runs in $\FRun(\control{C}{\Sigma})$ can be
    extended to an infinite run in $\IRun(\Sigma)$,
\item $\traj \vDash \Phi$ for any $\traj \in \Traj(\control{\C}{\Sigma})$, and
\item 
$\displaystyle
\liminf_{h \to \infty} \frac{1}{h} \sum_{i = 1}^ h \len(u_i) \geq \nu
$ for any $x_0 u_0 \ldots \in \IRun(\control{\C}{\Sigma})$.
\end{itemize}
\end{defn}
The first condition simply ensures that the controlled system will not
reach a deadlock, while the other two are the actual control
objectives.

\section{Problem Reduction to Mean-payoff Parity Games} \label{section: reduction} 

\subsection{Reduction to Symbolic Control} \label{section: symbolic control} 

In this section, we state our symbolic controller synthesis problem,
which considers a discrete system obtained by quantising states and
inputs, and by restricting control signals to piecewise-constant ones.
We show that a symbolic controller for this problem also satisfies the
conditions in Definition~\ref{defn: problem continuous}. 

A symbolic model is a tuple $(Q,Q_\ini,\VVV,\delta)$, where $Q$,
$Q_\ini$, and $\VVV$ are finite sets respectively of discrete states, initial
states, and signals, and $\delta \subseteq Q \times \VVV \times Q$ is
a transition function.
The notions of runs and finite runs carry directly to
symbolic models.
Similarly for controllers, which we call \emph{symbolic controllers}
in this case.

Given a time step $\tau\in \Rpos$, signal length bounds $\ell=
[\ell_\text{min}, \ell_\text{max}]$, and a discretisation parameter
$\mu \in \Rpos$, let
\begin{align*}
  \UUU_{\tau, \ell, \mu} =  \bigcup_{ j\tau \in \ell, j \in  \Zpos} \{ u: &[0, j\tau] \to U_\mu \} \mid 
  \forall i \in \{0, \ldots,j-1\}, 
  \\
  &
  \forall t \in [i \tau , (i+1) \tau),
  u(t) = u(i \tau) 
  \}\rlap{,}
\end{align*}
be a set of piecewise-constant control signals, where
\begin{equation}\label{eq: quantise}
U_\mu = \big\{\begin{bmatrix} u_1 & \cdots & u_m
	\end{bmatrix} ^\intercal \in U \bigm| 
	u_i = 2 \mu l_i ,\ l_i \in \Z,\ i \leq m \big\} 
\end{equation}
is the input set quantised by $m$-dimensional hypercubes of edge length $2\mu$.
Since $U$ is bounded, each signal $u \in \UUU_{\tau,
\ell, \mu}$ is a concatenation of constant signals of length $\tau$
and value in the finite input set $U_\mu$, so $\UUU_{\tau, \ell, \mu}$
is finite.

For a given $\eta \in \Rpos$, let
\begin{equation}\label{eq: quantise bigger}
  \Dis{X}{\eta} = \big\{x \in \mathbb{R}^n \bigm|
    x_i = 2 \eta l_i ,\ l_i \in \Z,\ i \leq n \,\wedge\,
    \Ball{x}{\eta}\cap X \neq \varnothing\big\}\rlap{.}
\end{equation}

\begin{defn}
\label{defn: symbolic model}
  The \emph{symbolic model} of a system $\Sigma$ for input-space
  quantisation parameters $\tau$, $\ell$, and $\mu$, and state-space
  quantisation parameter $\eta \in \Rpos$ is
\begin{equation*}
  \SSS_{\eta,\tau, \ell, \mu} =
    (Q = \Dis{X}{\eta}, Q_\ini = \Dis{X_\ini}{\eta},
    \UUU_{\tau, \ell, \mu}, \delta) 
\end{equation*} 
such that
$(q ,u, q') \in \delta$ if $(q ,u, q') \in Q \times \UUU_{\tau, \ell, \mu} \times Q$ and
\switchVersion{shortVersion}{
\begin{enumerate} 
\item $\forall x \in \Ball{q}{\eta}$ and $t \leq \len(u)$, $\xi_{x,u}^\fw(t) \subseteq X$, and
\item $\exists x' \in \xi_{q,u}^\rightarrow (\len(u))$,
	$
	 	\lVert x' - q' \rVert \leq \beta_u^\fw(\eta, \len(u)) + \eta, \text{ and}
	$
\item $\exists x \in \xi_{q',u}^\leftarrow (\len(u))$,
	$\lVert x - q \rVert \leq \beta_u^\bw(\eta, \len(u)) + \eta.
	$
\end{enumerate}}
{
\begin{enumerate}
\item for all $x \in \Ball{q}{\eta}$ and $t \leq \len(u)$, $\xi_{x,u}^\fw(t) \subseteq X$,
\item there exists $x' \in \xi_{q,u}^\rightarrow (\len(u))$ such that
	\begin{equation*}
	 	\lVert x' - q' \rVert \leq \beta_u^\fw(\eta, \len(u)) + \eta, \text{ and}
	\end{equation*} 
\item there exists $x \in \xi_{q',u}^\leftarrow (\len(u)),$ such that
	\begin{equation*}\lVert x - q \rVert \leq \beta_u^\bw(\eta, \len(u)) + \eta.
	\end{equation*} 
\end{enumerate}}
\end{defn} 

Notice that $\Dis{X}{\eta}$ may contain some points that are not in $X$,
but they have no outgoing transition in $\delta$, so they will not influence our controller synthesis algorithm.

\switchVersion{shortVersion}{}{
\begin{remark}\label{remark: delta_under_approx}
To compute the first condition of $\delta$, we may under-approximate it by 
$\Ball{q}{\alpha^\fw(\eta, \len(u))} \subseteq X$.
\end{remark}
}

\begin{remark}
  We can also use multi-dimensional quantisation parameters $\mu =
  \begin{bmatrix} {\mu}_1 & \cdots & {\mu}_m \end{bmatrix}^\intercal
  \in \Rpos^m$ and $\eta = \begin{bmatrix} {\eta}_1 & \cdots &
  {\eta}_n\end{bmatrix} ^\intercal \in \Rpos^n$.
  In this case, Equation~\eqref{eq: quantise} becomes
  \begin{equation*} 
  U_\mu = \big\{\begin{bmatrix} u_1 & \cdots & u_m
  	\end{bmatrix}^\intercal \in U \mid 
  	u_i = 2 \mu_i l_i ,\ l_i \in \Z,\ i \leq m \big\},
  \end{equation*}
  and similarly for Equation~\eqref{eq: quantise bigger}.
\end{remark}

A symbolic controller $S$ of $\SSS_{\eta,\tau,\ell,\mu}$ can directly
be turned into a controller $C_S$ of $\Sigma$ by $C_S(x_0 u_0 \ldots
x_k) = S(\overline{x_0} u_0 \ldots \overline{x_k})$, where
$\overline{x}$ is the closest point to $x$ in $Q$.

\begin{defn} \label{defn: problem symbolic}  
Given a system $\Sigma = (X, X_\ini, U, {\UUU}, \xi^\fw, \xi^\bw)$, a set $AP$ of atomic propositions, a function 
$P: X \to  \powerset(AP)$,
a $\logic$ path formula $\Phi$, 
 a threshold $\nu \in \Rpos$, and quantisation parameters $\tau, \ell, \mu, \eta$, 
the \emph{symbolic controller synthesis problem} consists in synthesising a symbolic controller
$S:\FRun(\SSS_{\eta,\tau, \ell, \mu}) \to \UUU_{\tau, \ell, \mu}$ 
such that 
\begin{itemize}
  \item all finite runs in $\FRun(\control{C_S}{\Sigma})$ can be
    extended to an infinite run in $\IRun(\Sigma)$,
\item $\traj \vDash \Phi$ for any $\traj \in \Traj(\control{\C_S}{\Sigma})$, and
\item 
$\displaystyle
\liminf_{h \to \infty} \frac{1}{h} \sum_{i = 1}^ h \len(u_i) \geq \nu
$ for any $x_0 u_0 \ldots \in \IRun(\control{\C_S}{\Sigma})$.
\end{itemize}
\end{defn}

\begin{thm}[extended from~\cite{icarcv20}]
\label{thm: problem_reduction}
If $S:\FRun(\SSS_{\eta,\tau, \ell, \mu})$ $\to \UUU_{\tau, \ell, \mu}$  solves the
problem of Definition~\ref{defn: problem symbolic},
then $\C_S: \FRun(\Sigma) \to \UUU_{\tau, \ell, \mu}$ 
solves the controller synthesis problem of Definition~\ref{defn: problem continuous}.
\end{thm}

The proof of this theorem can be found in~\cite{icarcv20} and relies
on the notion of \emph{alternating approximate simulation}, which
holds by Assumptions~\ref{asm: inc fw and bw complete}
and~\ref{asm: Lipchitz}.

\begin{remark}
  It may be that the problem in
  Definition~\ref{defn: problem symbolic} cannot be solved, but the
  one in Definition~\ref{defn: problem continuous} can be.
  Therefore, our approach is sound, but not complete.
\end{remark}


\subsection{Atomic Propositions along Symbolic Transitions}

We want to recover some information about visited states along
trajectories, which is lost in the symbolic model.
More precisely, we need to know which atomic propositions hold along
trajectories.
To this end, we introduce functions $\pmay,\pmust:\delta \to
\powerset(AP)^2$
that under-approximate these sets.
They will be crucial in the problem translation in
Section~\ref{subsec: problem trans}.

For all transitions $(q,u,q') \in \delta$ in the symbolic model $\SSS_{\eta,\tau, \ell, \mu}$, we require that
for $\Finv \in \ens{\forall, \exists}$,
\begin{align}
  & \text{if }\rho_\Finv(q,u,q') = (P^+,P^-) \text{ then }\nonumber\\
  \label{eq pmay}
  & \quad \forall x \in \Ball{q}{\eta},
    \forall x' \in \Ball{q'}{\eta},
    \forall \sigma \in \FTraj(\Sigma, x u x'),\\
  & \quad \Finv t \leq \len(u),
    P^+ \subseteq P(\sigma(t)) \wedge
    P^- \cap P(\sigma(t)) = \emptyset \rlap{.}\nonumber
\end{align}
The intuition is as follows. If $\rho_\forall(q,u,q') = (P^+,P^-)$ (\emph{resp.} $\rho_\exists(q,u,q')$),
then, along the transition $(q,u,q')$, each $p \in P^+$ holds at all
times (\emph{resp.} at some time) and no $p \in P^-$ holds at any time
(\emph{resp.} each $p \in P^-$ does not hold at some time).
Then, we can define $\rho_\Finv(q,u,q') \vDash \phi$ inductively on
the state formula $\phi$ as usual.

For the implementation, 
we use functions $B^+, B^-: X \times \Rpos \to \powerset(AP)$  such that, 
for all states $x \in X$ and radii $r \in \Rpos$, 
 $B^+(x,r) = \{ p \in AP \mid  \forall x' \in \Ball{x}{r}, x' \vDash p\}$
 and
 $B^-(x,r) = \{ p \in AP \mid  \forall x' \in \Ball{x}{r}, x' \nvDash p\}$ 
are the sets of atomic propositions
that are satisfied and not satisfied, respectively, at all states in the ball $\Ball{x}{r}$.
In the latter, we assume that, for all states $x \in X$ and radii $r
\in \Rpos$, $B^+(x,r)$ and $B^-(x,r)$ can be computed.

Then, 
we may use the following functions $\pmay$ and $\pmust$.
{\switchVersion{shortVersion}{\tinydisplayskip}{}
\begin{align*}
\pmay(q,u,q') = &\big(B^+(q,\eta) \cup B^+(q',\eta), B^-(q,\eta) \cup B^-(q',\eta)\big).\\
\pmust(q,u,q') = &\big(B^+(q,r) \cap B^+(q',r), B^-(q,r) \cap B^-(q',r)\big),\\
					&\text{ where } r =\beta^\fw_u(q, \len(u))+ \alpha^\fw_u(q,\len(u)). 
\end{align*}}
By Assumptions \ref{asm: inc fw and bw complete} and \ref{asm: Lipchitz}, $\pmay$ and $\pmust$
satisfy Equation~\eqref{eq pmay}.

\subsection{Mean-payoff Parity Games}
\label{subsec:mppg}
We recall some known results about \emph{\mppgs{} (MPPGs)} that
we use for solving the symbolic controller synthesis problem.
We invite the interested reader to see~\cite{chaterjee05} for more details about MPPGs.
\begin{defn}
  A \emph{\mppg{}} is a tuple {$\GGG = \big(G=(V = V_1 \sqcup V_2, E=(E_{1 \to 2} \sqcup E_{2 \to 1}), s
      \colon E \to V, t \colon E \to V), \lambda, c, \nu \big)$}, where 
  \begin{itemize}
    \item $G$ is a directed bipartite graph.
      	$V$ is partitioned into two disjoint sets $V_1$ and $V_2$ 
      	 of vertices for \emph{Player-1} and \emph{Player-2}, respectively.
      	 $E$ is its set of edges. Functions $s$ and $t$ map edges to their sources and targets.
    \item $\lambda \colon E \to \N$ maps each edge to its
      \emph{payoff}.
    \item $c \colon V \to \N$ maps each vertex to its \emph{colour}.
    \item $\nu \in \N$ is a given \emph{mean-payoff threshold}.
  \end{itemize}
\end{defn}

A \emph{play} on $\GGG$ is an infinite sequence 
$\omega = v_0 e_0 v_1 e_1 \ldots \in (VE)^\omega$ such that, for all
$i \geq 0$, $s(e_i) = v_i$ and $t(e_i) = v_{i+1}$.
A \emph{finite play} is a finite sequence in $V(EV)^*$ defined
similarly.
Let $\FPlay$ be the set of all finite plays, and $\FPlayi$ and
$\FPlayii$ be the set of those ending with a vertex in $V_1$ and $V_2$,
respectively.
Both players play according to strategies.
A strategy of Player-$i$ is a partial function
$\sigma_i \colon \FPlay_i \rightharpoonup E$ such that $s(\sigma_i(v_0 e_0 \ldots
 v_n)) = v_n$, i.e., $\sigma_i$ chooses an edge whose
source is the ending vertex of the play if such an edge exists,
and is undefined otherwise.
A play $\omega = v_0 e_0 v_1 e_1 \ldots$ is consistent with $\sigma_i$
if $e_j = \sigma_i(v_0 e_0 \ldots v_j)$ for all $v_j \in V_i$.
For an initial vertex $v$ and strategies $\sigma_1$ and $\sigma_2$ for
both players, we denote by $\mathit{play}(v, \sigma_1,\sigma_2)$ the
unique play consistent with both $\sigma_1$ and $\sigma_2$.
This play may be finite if a player cannot choose an edge.

For an infinite play $\omega = v_0 e_0 v_1 e_1 \ldots$,
we denote the maximal colour that appears infinitely often in the 
sequence $c(v_0) c(v_1) \ldots$ by $\Inf(\omega)$.
The \emph{mean-payoff value} of $\omega$ is  
$
\MP(\omega) = \liminf_{n \to \infty} \frac{1}{n} \sum_{i = 1}^n \lambda(e_i)
$.
A vertex $v \in V$ is \emph{winning} for Player-1 
if there exists a strategy
$\sigma_1$ of Player-1 such that,
for any strategy $\sigma_2$ of Player-2, 
$\mathit{play}(v, \sigma_1,\sigma_2)$ is infinite,
$\Inf(\mathit{play}(v, \sigma_1,\sigma_2))$ is even and 
\switchVersion{shortVersion}{$MP(\mathit{play}(v, \sigma_1,\sigma_2)) \geq \nu$.}{$MP(\mathit{play}(v, \sigma_1,\sigma_2))$ is greater than the mean-payoff threshold $\nu$.}
Such a $\sigma_1$ is called a \emph{winning strategy} for Player-1 from the vertex $v$.
The \emph{threshold problem} \cite{Daviaud2018pseudo} is 
to compute the set of winning vertices of Player-1 for a given MPPG.

In \cite{Daviaud2018pseudo}, the authors propose a pseudo-quasi-polynomial algorithm that solves the threshold
problem and computes a winning strategy for Player-1 from each winning state.
In Section \ref{subsec: problem trans}, 
we reduce the symbolic control problem to the synthesis of a winning strategy on an MPPG,
which can be solved using this algorithm.

\subsection{Problem Translation to Mean-payoff Parity Games}\label{subsec: problem trans}

  We translate the control problem for a formula
  $\Phi$ and a threshold $\nu$ to finding a winning strategy for a
  mean-payoff parity game $\GGG_{\Phi,\nu}$.
  All the constructions are given in~\cite{icarcv20} except for Until,
  so we only give intuitions here.

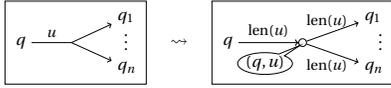
\begin{figure}[t]
    \begin{center}
    \resizebox{.3\textwidth}{!}{
      \begin{tikzpicture}
        \node[draw] (l) at (0,0) {
          \begin{tikzpicture}
            \coordinate (m) at (0,0);
            \node[anchor=east] (ll) at (-\basehlength,0) {$q$};
            \node[anchor=west] (r1) at ( \basehlength, \basevlength) {$q_1$};
            \node[anchor=west] (rn) at ( \basehlength,-\basevlength) {$q_n$};
            \node () at ($(r1.south)!0.5!(rn.north)$) {$\rvdots$};
            \path (ll) edge node[above] {\small$u$} (m)
                  (m)  edge[->] (r1)
                       edge[->] (rn);
          \end{tikzpicture}
        };
        \node[draw,anchor=west] (r) at ($(l.east)+(1.3,0)$) {
          \begin{tikzpicture}[
            onode/.style={circle,draw, inner sep=0,minimum size=0.5em},
          ]
            \node[anchor=east] (ll) at (0,0) {$q$};
            \node[onode] (m) at (1.5*\basehlength,0) {};
            \node[anchor=west] (r1) at (2*1.5*\basehlength, \basevlength) {$q_1$};
            \node[anchor=west] (rn) at (2*1.5*\basehlength,-\basevlength) {$q_n$};
            \node () at ($(r1)!0.5!(rn)$) {$\rvdots$};
            \node[ellipse callout,callout absolute pointer=(m.south west),draw,anchor=north east,inner sep=0] () at ($(m.south west)-(0.3,0.2)$) {$(q,u)$};
            \path[->] (ll) edge node[above] {\small$\len(u)$} (m)
                      (m) edge node[above] {\small$\len(u)\ \ $} (r1)
                          edge node[below] {\small$\len(u)\ \ $} (rn);
          \end{tikzpicture}
        };
        \node () at ($(l.east)!0.5!(r.west)$) {$\leadsto$};
      \end{tikzpicture}
      }
    \end{center}
    \caption{Construction of the base game}
    \label{fig:game:base}
  \end{figure}

  Let us start with the construction of the base game, illustrated in
  Fig.~\ref{fig:game:base}.
  There, Player-1 corresponds to the controller, and Player-2 to the
  environment.
  Each discrete state $q$ is mapped to a Player-1 vertex, and each
  pair $(q,u)$ is mapped to a Player-2 vertex.
  Each $q$ has an edge to $(q,u)$, which corresponds to the controller
  sending control signal $u$.
  Each $(q,u)$ has an edge to $q'$ if $(q,u,q') \in \delta$, which
  corresponds to resolving the environmental non-determinism.
  All such edges have payoff $\len(u)$, and the colouring of states is
  undefined (it is defined later by induction on the formula).

  Let us consider the discrete system in
  Fig.~\ref{fig:discrete_system}, which has a single input signal $u$,
  so we omit it for readability.
  The constructions of $\GGG_{\Globally p_2,\nu}$ and $\GGG_{\Repeatt
  p_1, \nu}$ are given in Fig.~\ref{fig:game:base cases} (the other
  cases are similar).
  Each of them contains two copies (coloured green and red in
  Fig.~\ref{fig:game:base cases}) of the base game.
  Colours are constant on each copy, and given by the coloured numbers
  in Fig.~\ref{fig:game:base cases}.
  Edges from $q$ to $(q,u)$ always stay in the same copy, while edges
  from $(q,u)$ to $q'$ may switch to a different copy (in
  Fig.~\ref{fig:game:base cases}, arrows from $(q,u)$ are coloured
  with the colour of the copy they point to).
  For $\Globally p_2$, an edge $(q,u) \to q'$ from the first copy
  points to the second one if $\pmust(q,u,q') \nvDash p_2$, and edges
  from the second copy always point there; the intuition is that the
  second copy is a losing copy for Player-1, and we should move to it
  if at some point it cannot be shown that $p_2$ holds all the time
  along the transition.
  For $\Repeatt p_1$, an edge $(q,u) \to q'$ points to the second copy
  if $\pmay(q,u,q') \vDash p_1$ and to the first one otherwise
  (independently of the starting copy); the intuition being that the
  second copy detects points where it can be shown that $p_1$ holds at
  some point along the transition, and it needs to be visited
  infinitely often to win the game.

  \begin{figure}[t]
    \centering
    \begin{minipage}{0.13\textwidth}
      \centering
      \begin{tikzpicture}[
        pnode/.style={
        },
        onode/.style={circle, draw, inner sep=0,minimum size=0.5em},
        show background rectangle,
      ]
        \node[pnode] (b) at (0,-\vlength/2) {$q'$};
        \node[pnode] (t) at (0, \vlength/2) {$q$};
        \path[->] (t) edge[bend right] node[left, align=center] {$p_1^\exists p_2^\forall$} (b)
                  (b) edge[bend right] node[right,align=center] {$p_2^\exists$} (t);
      \end{tikzpicture}
    \end{minipage}
    \hfil
    \begin{minipage}{0.25\textwidth}
      where $q \xto{u,p_1^\exists p_2^\forall} q'$ stands for\\
      $\left\{
        \begin{array}{l}
          \pmust(q,u,q') = (\ens{p_2},\emptyset) \\
          \pmay (q,u,q')  = (\ens{p_1,p_2},\emptyset)\rlap{,}
        \end{array}
      \right.$
      and similarly for $q' \xto{u,p_2^\exists} q$.
    \end{minipage}
    \caption{A simple discrete system}
    \label{fig:discrete_system}
  \end{figure}
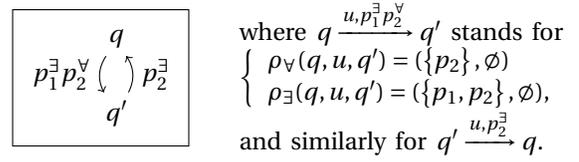

  \begin{figure}[t]
    \centering
    \begin{tikzpicture}[
      pnode/.style={
        inner sep=0
      },
      onode/.style={circle, draw, inner sep=0,minimum size=0.5em},
      show background rectangle,
    ]
      \node[pnode] (lb) at (0,-\vlength/2) {$q'$};
      \node[pnode] (lt) at (0, \vlength/2) {$q$};
      \node[onode] (ll) at (-\hlength/2,0) {};
      \node[onode] (lr) at ( \hlength/2,0) {};
      \node[draw=vert,fit=(lb) (lt) (ll) (lr)] (l) {};
      \node[pnode] (rb) at ($(2*\hlength,0)+(0,-\vlength/2)$) {$q'$};
      \node[pnode] (rt) at ($(2*\hlength,0)+(0, \vlength/2)$) {$q$};
      \node[onode] (rl) at ($(2*\hlength,0)+(-\hlength/2,0)$) {};
      \node[onode] (rr) at ($(2*\hlength,0)+( \hlength/2,0)$) {};
      \node[draw=rouge,fit=(rb) (rt) (rl) (rr)] (r) {};
      \node[anchor=south west,inner sep=0,align=left] at ($(l.north west)+(0,0.3333em)$)
        {$\GGG_{\Globally p_2,\nu}$};
      \node[vert,anchor=north west] at (l.north west) {$0$};
      \node[rouge,anchor=north west] at (r.north west) {$1$};
      \path[->] (lt) edge (ll)
                (lb) edge (lr)
                (rt) edge (rl)
                (rb) edge (rr)
                (rl) edge[rouge] (rb)
                (rr) edge[rouge] (rt)
                (ll) edge[rouge]  (rb)
                (lr) edge[vert] (lt);
    \end{tikzpicture}
    \hfil
    \begin{tikzpicture}[
      pnode/.style={
        inner sep=0
      },
      onode/.style={circle, draw, inner sep=0,minimum size=0.5em},
      show background rectangle,
    ]
      \node[pnode] (lb) at (0,-\vlength/2) {$q'$};
      \node[pnode] (lt) at (0, \vlength/2) {$q$};
      \node[onode] (ll) at (-\hlength/2,0) {};
      \node[onode] (lr) at ( \hlength/2,0) {};
      \node[draw=vert,fit=(lb) (lt) (ll) (lr)] (l) {};
      \node[pnode] (rb) at ($(2*\hlength,0)+(0,-\vlength/2)$) {$q'$};
      \node[pnode] (rt) at ($(2*\hlength,0)+(0, \vlength/2)$) {$q$};
      \node[onode] (rl) at ($(2*\hlength,0)+(-\hlength/2,0)$) {};
      \node[onode] (rr) at ($(2*\hlength,0)+( \hlength/2,0)$) {};
      \node[draw=rouge,fit=(rb) (rt) (rl) (rr)] (r) {};
      \node[anchor=south west,inner sep=0,align=left] at ($(l.north west)+(0,0.3333em)$)
        {$\GGG_{\Repeatt p_1,\nu}$};
      \node[vert,anchor=north west] at (l.north west) {$1$};
      \node[rouge,anchor=north west] at (r.north west) {$2$};
      \path[->] (lt) edge (ll)
                (lb) edge (lr)
                (rt) edge (rl)
                (rb) edge (rr)
                (ll) edge[rouge]  (rb)
                (rl) edge[rouge]  (rb)
                (lr) edge[vert] (lt)
                (rr) edge[vert] (lt);
    \end{tikzpicture}
    \caption{Construction of $\GGG_{\Globally p_2,\nu}$ and $\GGG_{\Repeatt p_1,\nu}$}
    \label{fig:game:base cases}
  \end{figure}

  We now explain the new case, namely the Until case.
  A graphical description of the construction of $\GGG_{p \Until
  \Phi, \nu}$ is described in Fig.~\ref{fig:game:until}.
  It contains $\GGG_{\Phi,\nu}$, which comes equipped with
  an \emph{initial} copy $I$ of the base game (for example, in the
  games of Fig.~\ref{fig:game:base cases}, $I$ is the green copy).
  It also contains a new copy of the base game.
  The intuition is that this new copy will be used to encode the
  verification of the $p$ part of $p \Until \Phi$, while
  $\GGG_{\Phi,\nu}$ will be used for $\Phi$.
  This new copy of the base game is different from the one shown in
  Fig.~\ref{fig:game:base} on two aspects.
  First, there are two Player-2 nodes for each pair $(q,u)$: edges
  from the first one stay in the new copy, while edges from the second
  one go to $I$.
  This corresponds to Player-1 making a choice whether to keep
  checking $p$ or to start checking $\Phi$.
  Second, there are edges from $q$ to a $(q,u)$ only if $p \in
  \pmust(q,u,q')$ for all $(q,u,q') \in \delta$.
  This is because $p$ must hold at all times
  regardless of system non-determinism in the first part of
  specification $p \Until \Phi$.
  In Fig.~\ref{fig:game:until}, the dashed edges from $q$ do not exist
  in the game, because one of the non-deterministic branches does not
  verify the condition above.

  \begin{figure}[t]
    \begin{subfigure}[b]{0.16\textwidth}
      \centering
      \scalebox{0.8}{
        \begin{tikzpicture}[
          show background rectangle,
          baseline=(q.base),
        ]
          \node[anchor=east] (q)   at (-\untilhlength,-3*\untilvlength/2) {$q$};
          \node[anchor=west] (q11) at (0,0)           {$q_1^1$};
          \node[anchor=west] (q12) at (0,-\untilvlength)   {$q_1^2$};
          \node[anchor=west] (q21) at (0,-2*\untilvlength) {$q_2^1$};
          \node[anchor=west] (q22) at (0,-3*\untilvlength) {$q_2^2$};
          \coordinate (u1)  at (-\untilhlength/2,-\untilvlength/2) {};
          \coordinate (u2)  at (-\untilhlength/2,-5*\untilvlength/2) {};
          \path (q) edge node [above] {\scriptsize $u_1\ \ $} (u1)
                    edge node [below] {\scriptsize $u_2\ \ $} (u2);
          \path[->] (u1) edge node [above] {\scriptsize $p^\forall$} (q11)
                         edge node [below] {\scriptsize $p^\forall$} (q12)
                    (u2) edge node [above] {\scriptsize $p^\forall$} (q21)
                         edge node [below] {\scriptsize $p^\exists$} (q22);
        \end{tikzpicture}
      }
      \caption{A discrete system}
    \end{subfigure}
    \begin{subfigure}[b]{0.35\textwidth}
      \centering
      \scalebox{0.7}{
        \begin{tikzpicture}[
          show background rectangle,
          onode/.style={circle, draw, inner sep=0,minimum size=0.5em},
        ]
            \node (rq11) at (0,0)           {$q_1^1$};
            \node (rq12) at (0,-\untilvlength)   {$q_1^2$};
            \node (rq21) at (0,-2*\untilvlength) {$q_2^1$};
            \node (rq22) at (0,-3*\untilvlength) {$q_2^2$};
            \node (Ildots) at (-\untilhlength/2,-3*\untilvlength/2) {\ldots};
            \node[draw,fit=(rq11) (rq12) (rq21) (rq22) (Ildots)] (I) {};
            \node[anchor=north west] (Ilab) at (I.north west) {$I$};
          \node[anchor=east] (spl) at ($(I.west)-(\untilhlength/2,0)$) {};
          \node[anchor=west] (spr) at ($(I.east)+(\untilhlength,0)$) {};
          \node[anchor=center] (subgldots) at ($(I.east)!0.67!(spr)$) {\ldots};
          \node[draw,fit=(I) (spl) (spr)] (right) {};
          \node[anchor=north west] (rightlab) at (right.north west) {$\GGG_{\Phi,\nu}$};
          \node (lq) at ($(right.west)-(3*\untilhlength,0)$) {$q$};
          \node (lq11) at ($(lq)+(2*\untilhlength,3*\untilvlength/2)$) {$q_1^1$};
          \node (lq12) at ($(lq)+(2*\untilhlength,\untilvlength/2)$) {$q_1^2$};
          \node (lq21) at ($(lq)+(2*\untilhlength,-\untilvlength/2)$) {$q_2^1$};
          \node (lq22) at ($(lq)+(2*\untilhlength,-3*\untilvlength/2)$) {$q_2^2$};
          \node[onode] (lqu1s) at ($(lq11)-(\untilhlength,0)$) {};
          \node[onode] (lqu1l) at ($(lq12)-(\untilhlength,0)$) {};
          \node[onode] (lqu2s) at ($(lq21)-(\untilhlength,0)$) {};
          \node[onode] (lqu2l) at ($(lq22)-(\untilhlength,0)$) {};
          \path[->] (lq) edge (lqu1s) edge (lqu1l)
                         edge[dashed] (lqu2s) edge[dashed] (lqu2l)
                    (lqu1s) edge (lq11) edge (lq12)
                    (lqu2s) edge (lq21) edge (lq22)
                    (lqu1l) edge[bend left=6] (rq11) edge[bend right=15] (rq12)
                    (lqu2l) edge[bend left=6] (rq21) edge[bend right=15] (rq22);
        \end{tikzpicture}
      }
      \caption{The corresponding game}
    \end{subfigure}
    \caption{Construction of $\GGG_{p \Until \Phi,\nu}$}
    \label{fig:game:until}
  \end{figure}

  Building $\GGG_{\Phi \land \Psi,\nu}$ and $\GGG_{\Phi \lor \Psi,\nu}$
  basically corresponds to synchronising parity automata by
  remembering, for each colour $c$ of the first automaton the largest
  colour $c'$ seen in the other since the last time $c$ was seen
  during the current execution.
  If $\Phi$ or $\Psi$ is an Until formula, the construction can be
  optimised to avoid state space explosion: one only needs to start
  remembering colours when both automata have finished checking the
  first part of the Until.

  \begin{thm}[extended from~\cite{icarcv20}]
    \label{thm:strat_to_control}
    From a winning strategy $\sigma$ for Player-1 in $\GGG_{\Phi,\nu}$,
    one can effectively compute a symbolic controller $C_\sigma$ for
    $\SSS_{\eta,\tau,\ell,\mu}$ that solves the symbolic
    controller synthesis problem of Definition~\ref{defn: problem symbolic}.
  \end{thm}

  The proof is an obvious extension of that in~\cite{icarcv20}.

\section{Controller Synthesis Algorithm}\label{sec: algo}

\subsection{Algorithm Overview}
The overview of our process is illustrated in Fig.~\ref{fig: flow}.
First, we discretise the system $\Sigma$ into the symbolic model $\SSS_{\eta,\tau, \ell, \mu}$ based on the quantisation parameters $\eta$, $\tau$, $\ell$, and $\mu$.
Then, using the heuristic pruning algorithm proposed in Section~\ref{section prune}, 
we disable the control signals that do not look promising to verify $\Phi$.
We transform the pruned symbolic model into a mean-payoff parity game, as discussed in Section~\ref{section: reduction}.
Then, we reduce the size of the mean-payoff parity game by removing the vertices that are not reachable from the initial state.
After solving the mean-payoff parity game,
if there exists a winning strategy for Player-1,
we translate it to a symbolic controller.
If the algorithm fails to compute a winning strategy,
we may refine the quantisation parameters (e.g., setting $\eta= \frac{\eta}{2}$, or $\mu= \frac{\mu}{2}$, or $\tau= \frac{\tau}{2}$)
and repeat the process until the parameters become smaller than a
predefined threshold.

A challenge faced in practice with discretisation is that
the generated systems and games are too large to solve for larger
state spaces.
We present both the heuristic pruning algorithm and the reachability
computation in the following subsections, and demonstrate their
effectiveness using the experimental results in
Section~\ref{section: illus example}.

\subsection{Heuristic Pruning}\label{section prune}
\begin{figure}
  \centering
    \begin{subfigure}[b]{0.28\textwidth}
      \centering
       \includegraphics[scale=0.5]{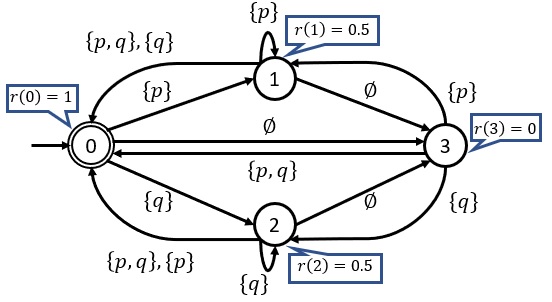}
      \caption{$\Box\Diamond p \wedge \Box\Diamond q$}
    \end{subfigure}
    \hfil
    \begin{subfigure}[b]{0.15\textwidth}
      \centering
       \includegraphics[scale=0.5]{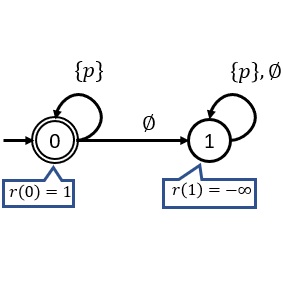}
      \caption{$\Box p$}
    \end{subfigure}
    \caption{B\"uchi automata with rewards assigned to all states.
	Each edge is labelled by a set of atomic propositions.}
    \label{fig: buchi}
  \end{figure}
\begin{figure}[t]
      \centering
      \includegraphics[scale=0.5]{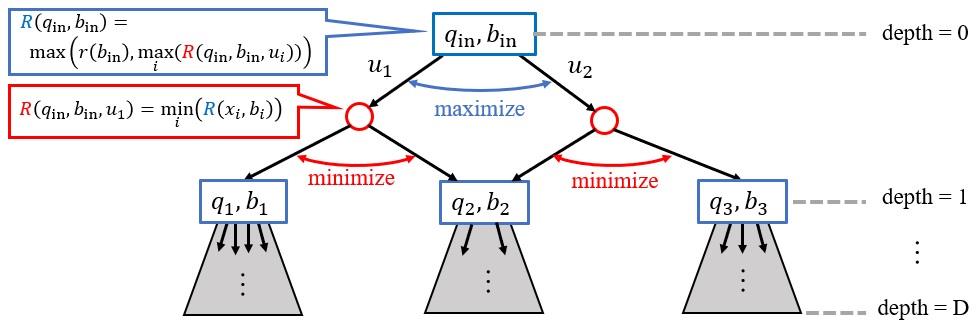}
      \switchVersion{shortVersion}{\squeezeupSmall}{}
      \caption{A finite-depth tree of the monitored runs used in the heuristic pruning algorithm.}
      \label{fig: prune_tree}
\switchVersion{shortVersion}{\squeezeupMid}{}
\end{figure}

We develop a heuristic pruning algorithm to only keep the control
signals that look most promising to verify $\Phi$.
For simplicity, we only describe the algorithm on a symbolic model
with a single initial state $q_\ini$.
We first translate $\Phi$ into its corresponding B\"uchi automaton $\BBB$,
which can be done using tools such as Spot~\cite{spot}.
Some examples of this translation are shown in Fig.~\ref{fig: buchi}. 
Then, we assign a reward to each B\"uchi state using a function $r: B \to \Rnonneg$, where $B$ is $\BBB$'s set of states, following the principles below.
For each accepting state $b \in B$ 
we assign $r(b) = 1$, which is the highest possible reward.
For each $b \in B$ from which it is impossible to reach any accepting state (e.g., state $1$ in Fig.~\ref{fig: buchi}(b)),
we assign $r(b) = -\infty$.
Otherwise, we may choose the reward $r(b)$ to be any value in $[0, 1)$.
For the example in Fig.~\ref{fig: buchi} (a), we assign reward 0.5 to the states $1$ and $2$, as they correspond to the case where one of the atomic proposition ($p$ or $q$) is detected.

Then, we monitor finite runs of length at most $D$ in the synchronised
product $\SSS_{\eta,\tau,\ell,\mu} \times \BBB$ from its initial state
$(q_\ini,b_\ini)$, based on which we disable some control signals. 
More precisely, we build a tree whose nodes are either in $Q \times B$
or in $Q \times B \times \UUU_{\tau,\ell,\mu}$, of depth $D$ as in
Fig.~\ref{fig: prune_tree}.
For each node $n$ at depth $d$ of the tree, we compute the estimated
reward $R(n)$ of $n$ as follows.
\begin{align}\label{eqV}
R(q,b) &=
\begin{cases}
0 &\text{if } d = D \\
  \displaystyle \max\big(r(b), \max_{u \in \UUU_{\tau, \ell, \mu}} R(q,b,u)\big) &\text{otherwise,}
\end{cases}
\end{align}
and
$\displaystyle
R(q,b,u) = \displaystyle
\min_{  \substack{(q, u, q') \in \delta,~ 
\rho_\exists(q, u, q') = (P^+, P^-),\\
(b,P^+,b')\text{ is a transition in }\BBB} }R(q',b')
$.

In words, $R(q_\ini,b_\ini)$ is the maximum reward $r(b)$ that the
controller can ensure to see in $\BBB$ for runs of length at most $D$.
If $R(q_\ini,b_\ini) = 1$, it means the system can be controlled to go
through an accepting state.
Red nodes in Fig.~\ref{fig: prune_tree} represent the non-determinism
of the system, which can go to any $(q', b')$ reached by $u$, so $R(q,b,u)$
has to be defined as a minimum of their expected rewards.

We prune the symbolic model at state $q$ by disabling signals $u$ that
do not maximise $R(q,b,u)$.
We also remove a state $q$ from $Q$ if all control signals are disabled at $q$.

Notice that each pair $(q,b) \in Q \times B$ may appear multiple times in the tree in Fig.~\ref{fig: prune_tree} (e.g., if there is a cyclic run).
To save computation time, we avoid computing $R(q,b)$ if $(q,b)$ is detected at depth $d$ 
and the value $R(q,b)$ has already been previously computed at depth $d' \leq d$.
As a result, our pruning algorithm is non-deterministic, depending on which branch of the tree we compute first. 
Also note that our pruning algorithm disables the signals for discrete
states, which correspond to several vertices in $\GGG_{\Phi,\nu}$.
Thus, there is a possibility that the algorithm prunes control signals 
that are needed for the controller to win.

Note that, in general, there are more than one initial state in
$Q_\ini$, in which case the algorithm extends directly using a forest
rather than a tree.

\begin{remark}
  \label{rem:pruning}
  Since the pruning process only prunes signals (and not the
  non-determinism), it only constrains the system, so a controller
  that solves the problem in Definition~\ref{defn: problem symbolic}
  for the pruned system also does it for the whole system.
\end{remark}

\subsection{Reachable Subgame}\label{section reach}
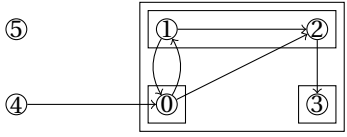
\begin{figure}[t]
	\centering
      \begin{tikzpicture}[
        pnode/.style={circle, draw, inner sep=0,minimum size=0.5em},
        onode/.style={circle, draw, inner sep=0,minimum size=0.5em},
      ]
        \node[pnode] (0) at (0,0) {$0$};
        \node[pnode] (1) at (0, \vlength) {$1$};
        \node[pnode] (4) at (-2*\hlength,0) {$4$};
        \node[pnode] (5) at (-2*\hlength,\vlength) {$5$};
        \node[pnode] (2) at (2*\hlength,\vlength) {$2$};
        \node[pnode] (3) at (2*\hlength,0) {$3$};
        \node[draw=black,fit=(0)] (w0) {};
        \node[draw=black,fit=(2) (1)] (w1) {};
        \node[draw=black,fit=(3)] (w2) {};
        \node[draw=black,fit=(w0) (w1) (w2)] (all) {};
        \path[->] (0) edge[bend right] (1)
                  (1) edge[bend right] (0)
                  (1) edge (2)
                  (0) edge (2)
                  (2) edge (3)
                  (4) edge (0);
      \end{tikzpicture}
      \caption{Reachable vertices from $W_0 = V_\ini = \{0\}$,
      	$W_1 = \{1,2\}$, $W_2 = \{3\}$, $W_3 = \varnothing$.}
      \label{fig:reach}
\end{figure}
We compute the reachable subgame of $\GGG_{\Phi, \nu}$ from initial
vertices $v_\ini \in V_\ini$ in a breadth-first traversal manner,
where $V_\ini$ is the set of initial states $q_\ini \in Q_\ini$ that
belong to the initial copy $I$ of $\GGG_{\Phi,\nu}$.
More precisely, we first set 
$W_0 = V_\ini$, and repeatedly compute the set $W_i$ of reachable vertices
from $V_\ini$ after exactly $i$ transitions. Concretely, we compute
$W_{i+1}$ as the set 
$\{v \in V\setminus\bigcup_{j \leq i}W_j \mid \exists e \in E, s(e) 
	\in W_i, t(e) = v\}$ 
until $W_{k} = \varnothing$ for some $k$.
Then, we apply a mean-payoff parity game solver on the subgame that 
contains the reachable vertices, i.e., $\bigcup_{j \leq k-1}W_j$. 
Fig.~\ref{fig:reach} shows an example of computation of the reachable 
subgraph. Observe that according to the definition, each vertex is visited 
at most once.

This technique may look simple, but it is already very efficient. 
Indeed, as we will see in Section~\ref{section: illus example}, this  
removes a large number of vertices. Actually, this allows to remove entire 
copies (as described in Section~\ref{subsec: problem trans}) in the game.

\begin{remark}
  \label{rem:reachability}
  Because the existence of a winning strategy is only affected by the
  reachable part of the game, each winning strategy on the reachable
  subgame is also a winning strategy on the whole game.
\end{remark}

\section{Experimental Results}
\label{section: illus example}

\begin{table}
\begin{center}
\footnotesize
\begin{tabular}{ |c|c|c|c|c|c|c|c| }
 \hline
  & no &  & prune & no &  & prune \\ 
 spec.& pre- & reach & + & pre- & reach & +\\
 &comp. & & reach & comp. &  &reach\\
 \hline
 \textbf{loop} & 202 & 36 & 33 & 82272 & 12866 & 4920\\
 \hline
 \textbf{2-loop} & 16131 & 853 & 157 & 658176 & 33780 & 4497\\
 \hline
 \textbf{until-1} & timeout & 2351 & 211${}^{\ast}$ & 709506 & 35107 & 12853 \\
 \hline
 \textbf{until-2} & timeout & timeout & 964 & 1181964 & 69744 & 39339 \\
 \hline
\end{tabular}
\caption{Computation times (seconds, columns 2-4) and size 
(in number of vertices, columns 5-7) of the games fed to the solvers. 
$\ast$ means no winning strategy is found.}
\label{tab:results}
\end{center}
\end{table}

As in \cite{icarcv20}, we consider a non-deterministic nonholonomic 
robot system. This is a modified version of \cite{SEMGL2019}, to allow 
non-determinism, coming from uncertainties in the measure of the velocity.
{\switchVersion{shortVersion}{\tinydisplayskip}{}
\begin{align*} 
\dot{x}(t) &= (1 + \lambda(t)) v  \cos(\theta(t)) &\\
\dot{y}(t) &= (1 + \lambda(t)) v  \sin(\theta(t)) 
& \dot{\theta}(t) = \omega(t)\rlap{,}
\end{align*}}
In this system, the input signal is given by $\omega$, the steering angle.
The physical dimensions are $x$, $y$, and $\theta$, respectively the
cartesian coordinates and heading angle.
The speed of the robot is $v$.
The non-determinism is given by $\lambda$, randomly chosen from $[- \bar{\lambda}, \bar{\lambda}]$ for $\bar{\lambda} \in \Rnonneg$. 

Functions $\beta^\fw$ and $\alpha^\fw$, as well as their 
backward versions, can easily be computed (more details are given in 
\cite{icarcv20}). We use our controller synthesis algorithm with the following 
parameters: 
$v = 1.5$, $\bar{\lambda} = 0.03$,
$X_\ini = \{(x,y,\theta) = (-5, -5, 0)\}$, 
$U = [-\frac{\pi}{2}, \frac{\pi}{2}]$, 
$\eta = \begin{bmatrix} 1 & 1 & \frac{\pi}{8} \end{bmatrix} ^ \intercal$,
$\mu = \frac{\pi}{2}$,
$\tau = \ell_\text{min} = 1$,
$\ell_\text{max} = 2$, and
$D = 10$.
The state space $X$ depends on the specification.
For the mean-payoff specification, we set the threshold $\nu = 1.5$, 
meaning that at least half of the input signals must be of length $2$.

\begin{figure}[t]
      \begin{subfigure}{.22\textwidth}
      \includegraphics[scale=0.24]{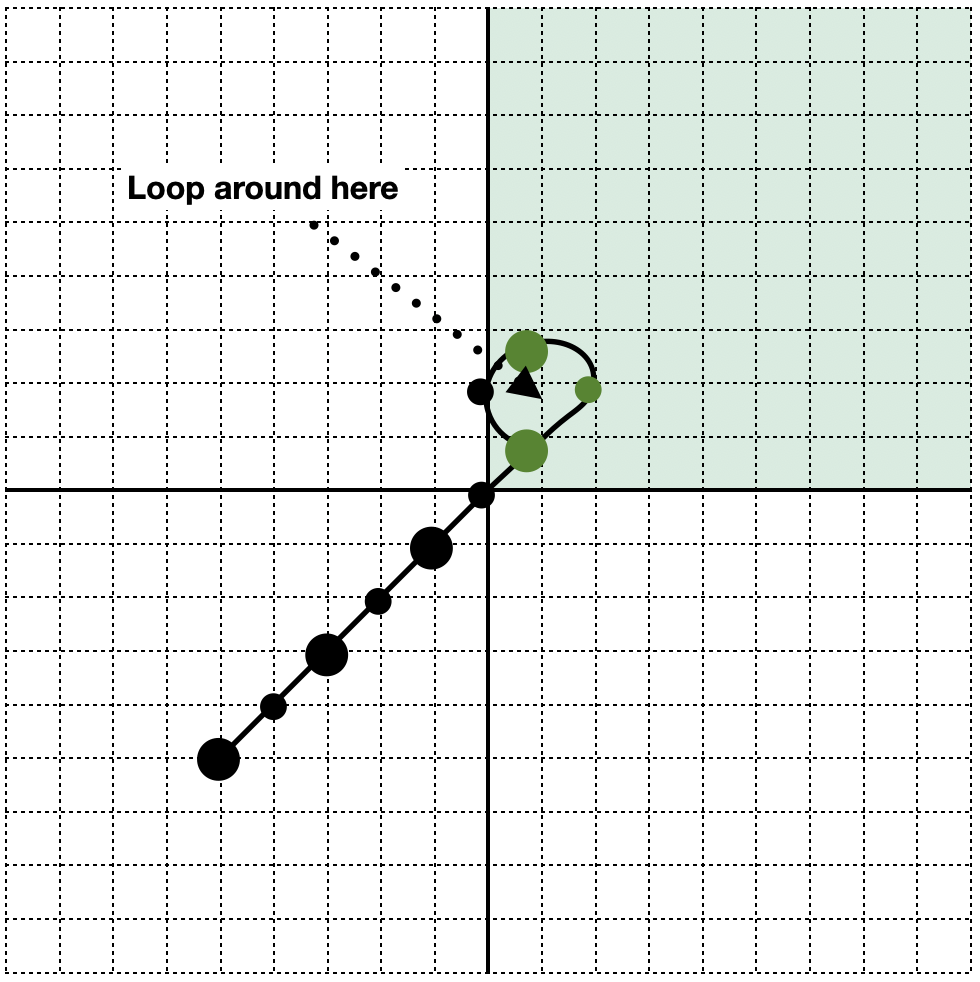}
        \caption{\textbf{loop}}
  	\label{fig:loop}
      \end{subfigure}
      ~
      \begin{subfigure}{.22\textwidth}
      \includegraphics[scale=0.24]{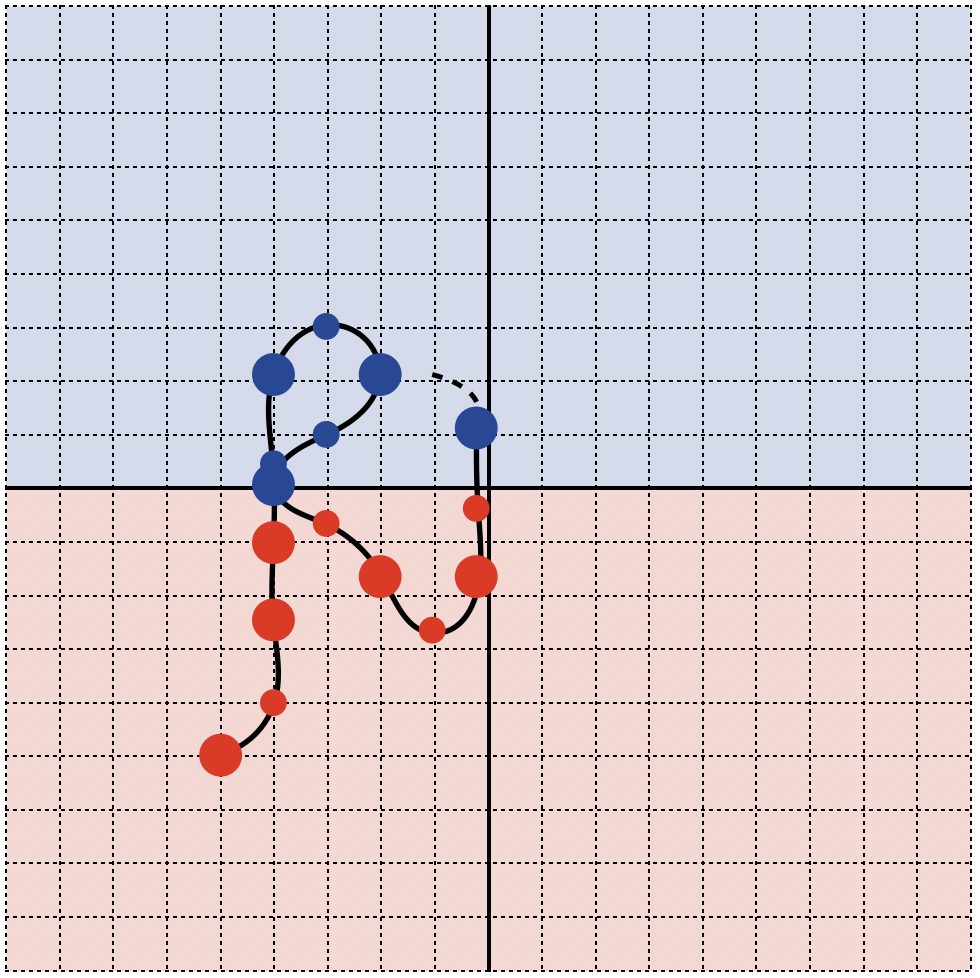}
        \caption{\textbf{2-loop}}
  	\label{fig:2-loop}
      \end{subfigure}
      
      \begin{subfigure}{.22\textwidth}
      \includegraphics[scale=0.24]{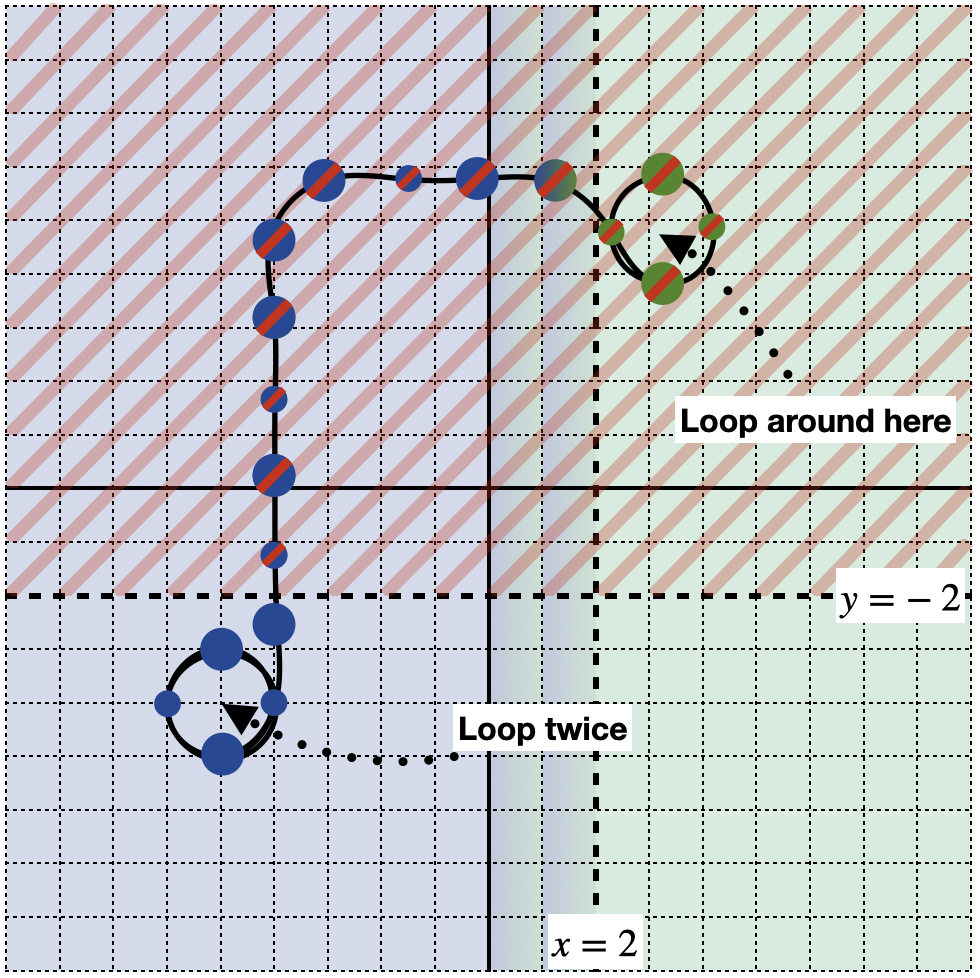}
        \caption{\textbf{until-1}}
  	\label{fig:until-1}
      \end{subfigure}
      ~
      \begin{subfigure}{.22\textwidth}
      \includegraphics[scale=0.24]{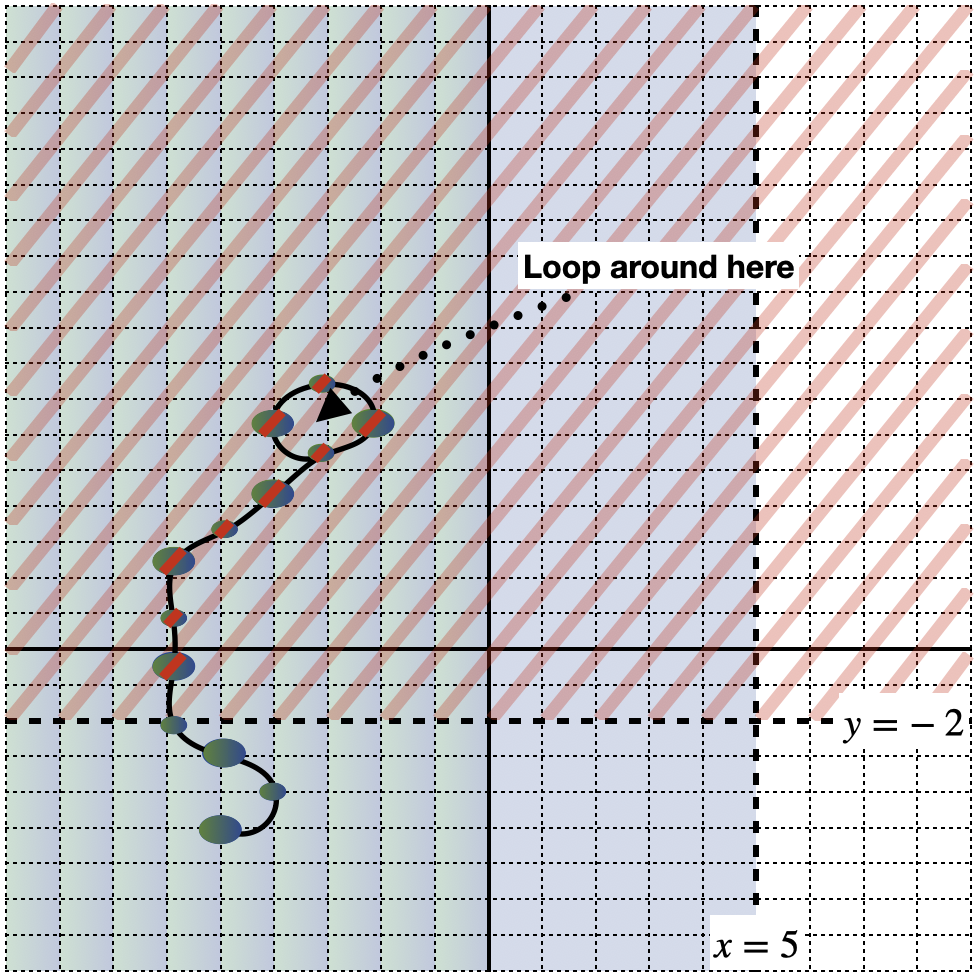}
        \caption{\textbf{until-2}}
  	\label{fig:until-2}
      \end{subfigure}
      \switchVersion{shortVersion}{\squeezeupSmall}{}
      \caption{A finite run under the synthesised controller for each specification.
      The bullets show the state of the robot at each time step $k\tau$.
      The bigger bullets represent the time steps where a control input is sent.
      The colours witness which atomic propositions hold at each time step.}
      \label{fig: robot}
      \switchVersion{shortVersion}{\squeezeupMid}{}
\end{figure}

We consider the four specifications depicted in Fig.~\ref{fig: robot}.
The \textbf{loop} specification (Fig.~\ref{fig:loop}) is given by the 
right-recursive LTL formula $\Box\Diamond\textbf{green}$, where 
$\textbf{green}$ stands for the atomic proposition $x > 0 \wedge y > 0$. 
This means that the robot must visit the green area infinitely often. 
The \textbf{2-loop} specification (Fig.~\ref{fig:2-loop}) is given by 
$\Box\Diamond\textbf{blue} \wedge \Box\Diamond\textbf{red}$, 
where $\textbf{blue}$ stands for $y > 0$, and $\textbf{red}$ for $y < 0$. 
In this case, the robot must navigate infinitely often between the upper and the lower 
parts of the state space.
The \textbf{until-1} specification (Fig.~\ref{fig:until-1}) is given by
$\textbf{blue}\Until(\Box\textbf{red}\wedge\Box\Diamond\textbf{green})$, 
where $\textbf{blue}$ is $x < 2$, $\textbf{red}$ is $y > -2$ and 
\textbf{green} is $x > 0$. Here, the robot must stay in the left side of the 
state space until it reaches and stays forever in the upper part, and 
it must visit the right side infinitely often.
These three specifications share the same state space 
$X = [-9,9]\times[-9,9]\times[0,2\pi]$.
Finally, the \textbf{until-2} specification (Fig.~\ref{fig:until-2}) is the same as 
\textbf{until-1}, except that $\textbf{blue}$ stands for $x < 5$, $\textbf{green}$ for $x < 0$ and 
the state space is larger $X = [-9,9]\times[-9,18]\times[0,2\pi]$.
Notice the increasing complexity in the specifications:
\textbf{2-loop} is more complex than \textbf{loop},
\textbf{until-1} than \textbf{2-loop}, and the state space of
\textbf{until-2} is larger than that of \textbf{until-1}.

To solve the mean-payoff parity game, we combine the reduction of
mean-payoff parity games to energy games in~\cite{chatterjee12} with
the solver for energy games in~\cite{brim10}.
The program was implemented in Python~3.8.6 and run on a MacBook Pro (Apple M1 chip, 16GB memory).
The results are compiled in Tab.~\ref{tab:results}. 
For each specification, we ran our algorithm without any 
precomputation, with reachability only, and with both reachability and 
heuristic pruning.
The times given for the cases using pruning are averaged
over $5$ executions, as this heuristic is non-deterministic.
The system studied in~\cite{icarcv20} is the \textbf{loop}
specification without any precomputation.

We observe that 
precomputations decrease both the size of the game, and the execution time.
The algorithm without precomputation easily reaches timeout 
(set at $5$ hours) when the specification becomes more 
complex. For \textbf{2-loop} and \textbf{until-2}, we observe that 
pruning makes the execution significantly faster, compared to reachability only. 
For \textbf{until-2}, reachability only is not even enough to avoid a timeout. 
There are two main reasons: first the size of the game is much larger; 
second, there are many more non-winning states for the mean-payoff 
specification, which makes the energy game solver (which uses value
iteration) much slower.

Finally, the third specification \textbf{until-1} witnesses the
limitations of pruning (already mentioned in
Section~\ref{section prune}): by pruning, we may remove some winning
strategies.
In this particular case, we remove all of them.
However, because our algorithm reaches this conclusion faster than
with reachability only, there is little harm in pruning.
 
\section{Conclusion and Future Work}
\label{section: conclusion}

We proposed a symbolic self-triggered controller synthesis
algorithm for non-deterministic continuous-time nonlinear systems without stability assumptions under two control specifications: a $\logic$ specification and a threshold for the average control signal length.
The main steps of the process are
1) to discretise the state and input spaces to obtain a symbolic model
corresponding to the original continuous system
2) to reduce the controller synthesis problem to the computation of a
winning strategy in a mean-payoff parity game.
In addition, we proposed a heuristic
pruning algorithm to speed up the computation by disabling some control
signals based on expected rewards in a B\"uchi automaton generated from the specification.
We demonstrated the efficiency of our method on the example of a nonholonomic robot
navigating in an arena under several specifications. 

For future work, we want to further investigate heuristics that help solve games in
practice by trying different variants and tradeoffs for our pruning algorithm.
One possibility would be to prune the game
-- rather than the symbolic model, which would be harder but would also
retain more strategies and could be done while computing the reachable set.
Another would be to prune from different states -- rather than only
from the initial states -- and see if it can improve performance.
Another direction is to explore different reward strategies for the
B\"uchi automata used by the heuristic.
Moreover, we want to develop a theory of B\"uchi automata with structured
alphabets that is suitable for our use.




%

\end{document}